\documentstyle[psfig]{l-aa}

\begin{document}

\title{LonGSp: the Gornergrat Longslit Infrared Spectrometer }

\author{L.~Vanzi\inst{1}, M.~Sozzi\inst{2}, G.~Marcucci\inst{1}, 
A.~Marconi\inst{1}, F.~Mannucci\inst{2}, F.~Lisi\inst{3}, L.~Hunt\inst{2}, 
E.~Giani\inst{1}, S.~Gennari\inst{3}, V.~Biliotti\inst{3}, and C.~Baffa\inst{3} }

\institute{
Dipartimento di Astronomia e Scienza dello Spazio, Universit\`a di Firenze,
Largo E. Fermi 5, I--50125 Firenze, Italy
\and
Centro per l'Astronomia Infrarossa e lo Studio del Mezzo Interstellare--CNR, 
Largo E. Fermi 5, I--50125 Firenze, Italy
\and
Osservatorio Astrofisico di Arcetri, Largo E. Fermi 5, 
I--50125 Firenze, Italy
}

\offprints{L. Vanzi}

\date{Received ...}

\thesaurus{ }
\maketitle
\markboth{Vanzi et al. }{}

\begin{abstract}
We present a near-infrared cooled grating spectrometer that has been
developed at the Arcetri Astrophysical Observatory for the 1.5 m 
Infrared Telescope at Gornergrat (TIRGO). 

The spectrometer is equipped with cooled reflective optics and a grating 
in Littrow configuration. The detector is an engineering grade Rockwell 
NICMOS3 array (256$\times$256 pixels of $40 \mu$m).
The scale on the focal plane is 1.73~arcsec/pixel and the field of view along
the slit is 70~arcsec.
The accessible spectral range is $0.95-2.5\mu$m with a dispersion, at 
first order, of about 11.5 {\rm\AA}/pixel.
This paper presents a complete description of the instrument, 
including its optics and cryo-mechanical system,
along with astronomical results from test 
observations, started in 1994.  Since January 1996, LonGSp is offered 
to TIRGO users and employed in several Galactic and extragalactic programs.


\end{abstract}

\section{Introduction}

The development of the spectrometer LonGSp (Longslit Gornergrat
Spectrometer) was part of a project aimed at providing 
the 1.5-m Infrared Telescope at Gornergrat (TIRGO)
with a new series of instruments based on 
NICMOS3 detectors. The infrared (IR) camera, ARNICA, developed in the context
of this project is described in Lisi et al. (1995). 
LonGSp is an upgrade of GoSpec (Lisi et al. 1990), the IR spectrometer
operating at TIRGO since October 1988. 
Thanks to the NICMOS detector, the new spectrometer
enables longslit spectroscopy with background
limited performance (BLP). The GoSpec characteristics of compactness and
simplicity are maintained in the new instrument.
Only a subsection of the engineering grade array (40$\times$256 pixels) is used.
A description of the optics, cryogenics, and 
mechanics is presented in Section \ref{opt} and \ref{cry}; the electronics,
software and the performance of the detector are presented in Section
\ref{elect} and \ref{detec}. Finally, in Section \ref{obs} we present 
details regarding the observations and data reduction, and in Section
\ref{astro} the results of the first tests at the telescope.

\section{Optical Design}
\label{opt}

The optical scheme of the instrument is sketched in Fig. 1; it is
designed to match the f/20 focal ratio of the TIRGO telescope. 

\begin{figure}
\centerline{\psfig{figure=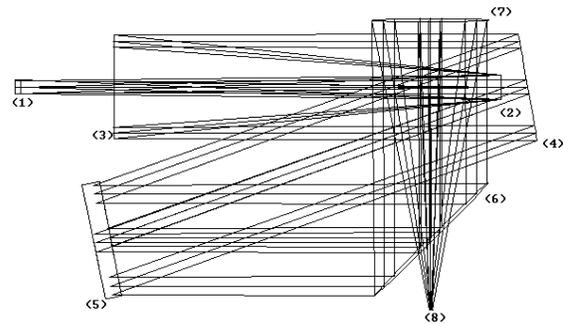,width=7.3cm,angle=-90}}
\caption{ 
Optical diagram of instrument.
The optical components of the instrument include (enumeration 
follows the path of radiation): (1) field lens, (2)
secondary mirror and (3) primary mirror of the collimator (an inverted
cassegrain), (4) a plane mirror, (5) the grating, (6) the plane mirror,
(7) the paraboloidal mirror of the camera, and (8) the detector.
}
\end{figure}

Following the optical path from the telescope,
the beam encounters the window of the 
dewar, the order sorting filter, a field lens, and the slit; the latter 
resides at the focal plane of telescope. 
The window and field lens are composed of calcium fluoride. 
Filters and slits are respectively mounted on two wheels and can be quickly 
changed during the observations.  The field lens images the pupil on
the secondary mirror of an inverted cassegrain (with focal length of
1400 mm) that produces a parallel beam 70 mm in diameter. This
beam is reflected onto the grating by a flat mirror tilted by $10^{\circ}$. 
The grating, arranged in Littrow configuration, has 150
grooves/mm and a blaze wavelength of $2 \mu$m at first order; rotation 
around the $10^{\circ}$ tilted axis allows the selection
of wavelengths and orders. 
A modified Pfund camera (with focal length of 225 mm) following the grating, 
collects the dispersed beams on the detector. The sky-projected pixel 
size is 1.73~arcsec, and the total field covered along the slit direction 
is 70~arcsec. 

The back face of the grating is a
flat mirror so that, when the grating is rotated by 180 degrees, the 
instrument functions as a camera,
in the band defined by the filters, 
with a field of view of about 1.5~arcmin square.
This facility can be useful for tests, 
maintenance, and for centering weak sources on the slit.

All the mirrors are gold coated to provide good efficiency over a
wide spectral range, and the optics are acromatic at least up to 
$5\mu$m. The optical components are cooled to about 80 K by means of 
thermal contact with a cryogenic vessel filled with liquid nitrogen 
at atmospheric pressure as described below. The mounting of optical 
elements is designed to take into account the dimensional changes 
between mirrors (in pyrex) and supports (in aluminium) generated by 
the cooling and the differences in thermal expansion coefficents. 

The resolving power is (for first order) about 600 
in the center of J band, and 950 in the center of the K band, using a 
slit of two pixels (3.46 arcsec).

\section{Cryogenics and Mechanics}
\label{cry}

As can be seen in Fig. 2, where we show some parts of the mechanical 
structure of the instrument,
the core of the instrument is the liquid nitrogen
reservoir, which has a toroidal shape with rectangular cross section and a
capacity of 3 liters. It
provides support and cooling for two optical benches, which are
located on opposite sides of the vessel.
The central hole of the toroid allows the beam to pass from one 
optical bench to the other.

\begin{figure}
\centerline{\psfig{figure=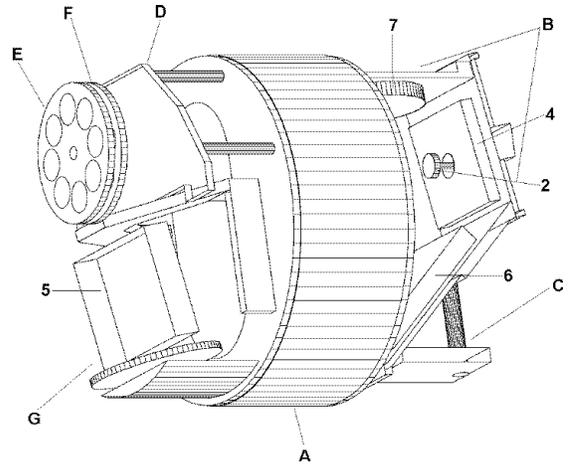,width=7.3cm,angle=-90}}
\caption{ 
Mechanical arrangement of instrument. 
The optical parts have the same enumeration as in Fig. 1. 
(A) liquid nitrogen vessel,
(B) support of mirrors in the back side of instruments,
(C) support of the detector,
(D) support of primary mirror of collimator,
(E) filter wheel,
(F) slit wheel (each slit support a field lens),
(G) grating mounting and wheel for positioning in wavelength.
}
\end{figure}

The grating motion is assured by an external stepper motor via a ferrofluidic
feedthrough, and the position is controlled by an encoder
connected to the motor axis outside the dewar.
Two springs acting on the worm gear guarantee good stability of the 
grating position.
Two internal stepper motors, modified to operate at cryogenic temperatures
(Gennari et al. 1993), drive the filter and slit wheels. 

Mechanics and optics are enclosed in a radiation shield. 
The internal cold structure is supported by nine low-thermal-conductivity
rods, which are fixed between the internal liquid nitrogen reservoir and
the external vacuum shield, and are rigidly linked to the focal
plane adaptor of the telescope. 
Externally, the instrument has the form of a cylinder with a base of 
about 40 cm in diameter and length of about 60 cm.

A small amount of active charcoal is present to maintain the
value of the pressure required (less than 10$^{-4}$ mb) for a
sufficiently long time (more than 20 days).
The charcoal is cooled by an independent cryogenic system; in the rear
optical bench there is a smaller nitrogen vessel ($\sim$ 0.5 l)
thermally insulated from the surrounding environment.
The regeneration of the charcoal must be carried
out once a month in order to maintain a sufficiently high absorption rate. 
This operation consists of heating the charcoal to 300 K, while
the pressure inside the dewar is maintained below 10$^{-1}$ mb by means
of a rotary vacuum pump. 
Because the charcoal is cooled by an
independent cryogenic system, the heating of the optics and the 
main part of the mechanical structure
is not necessary and the operation can be completed in about four hours.

To cool and warm the entire instrument reasonably quickly, the dewar is
filled with gaseous nitrogen at a pressure of about 200 mb during the cooling 
and heating phases: in this way the thermal transients prove to be shorter 
than seven hours. The rate of
evaporation of the nitrogen from the main reservoir
allows about 16 hours of operation in working conditions, more than
a winter night of observation.

\section{Electronics and Software}
\label{elect}

The electronics of LonGSp comprise two main parts: ``upper'' electronics,
that are situated near the instrument, and the ``lower'' electronics in the
control room. The connection between the two parts is assured 
via a fiber-optics link.
Two boards, close to the cryostat, house part of the interface electronics, 
that is a set of four preamplifiers and level shifters and an array of drivers 
and filters that feed the clock signals and the bias to the array multiplexer. 

The ``upper'' electronics are composed of an 
intelligent multi--part sequence
generator and a data acquisition section. The first is
controlled by a microprocessor (a Rockwell 65C02), and 
the sequencer is capable of generating many different waveforms template
(at 8 bits depth) stored in an array of 128~Kbytes of memory. The
final waveform is generated by selecting, via software, the templates needed
together with their repetitions. 

The data acquisition segment consists of a bank of four
analog-to-digital converters at 16 bits,
and the logic for converting them to serial format.  
A transceiver sends data to the telescope control room through a
fiber optics link. Data are sent as
groups of four, one for each quadrant, and are presented together with
the quadrant identifier (two bits) to the frame grabber.
The fiber optics link is bidirectional, so that it is possible to send
instructions to the control microprocessor in the ``upper'' electronics, and 
to communicate with the motor control through a serial connection (RS-232) 
encoded on the same fiber-optics link. The ``upper'' 
electronics are completed by the box which contains the power supply, the 
stepper motors controllers, and the temperature controller of the array.

The ``lower'' electronics implement the logic to decode the
serial data protocol, 
in order to correctly reconstruct the frame coming from the array detector.  
Data are collected by the custom frame grabber (known as the ``PingPong'') 
which 
is capable of acquiring up to four images in each of its two banks. When a 
bank is written, the other can be read, enabling continuous fast
acquisitions. Also of note is the ability to re-synchronize data
acquisition to the quadrant address, virtually eliminating mis--aligned
frames.  

The instrument is controlled by an MS-DOS PC equipped with a 80486 
CPU (33 MHz clock), high-resolution monitor, and 600 Mbytes of hard 
disk space. At the end of a data acquisition sequence, each single 
frame or the stack average of a group of frames is stored on 
the PC hard disk, and are later transferred to optical disk (WORM) storage.
In the near future, the WORM cartridges will be superseded by 
more standard writable CD-ROM cartridges. A local Ethernet network
connects the control computer to the TIRGO Sun workstation, 
so that it is possible to transfer the data for preliminary reduction
using standard packages.

The software developed for this instrument is ``layer organized'',
that is to say organized as a stack of
many layers of subroutines of similar levels of complexity. 
To accomplish its task, each routine
need rely only on the immediately adjacent level and on global utility
packages. 
Such a structure greatly simplifies the
development and maintenance of the software.

Our efforts were directed towards several different requirements. Our first
priority was to have a flexible laboratory and telescope engine,
capable of acquiring easily the large quantity of data a panoramic IR
array can produce. The human interface is realized through a 
fast character-based menu interface. The operator is presented only with the
options which are currently selectable, and the menu is rearranged on the
basis of user choices or operations. 

We have also stressed the auto--documentation of data. After the decision to 
store data in standard FITS format, it was deemed useful to fully 
exploit the header 
facility to label each frame with all relevant information, such as telescope
status, instrument status, and user acquisition choices. Data are also labelled 
with the observer name in order to facilitate data retrieval from our 
permanent archive.
In particular,
the form of the FITS file is completely compatible with the
context IRSPEC of the ESO package MIDAS. 

Finally, one of our main goals was to produce an easy-to-use software and
with the smallest ``learning curve''. Our idea was that data acquisition
must {\sl disappear} from observer attention, giving him/her the
possibility to concentrate on the details of the observations; 
in this way, observing efficiency is much higher.
As a result, we have implemented automatic procedures such as 
multi--position (``mosaic''), and multi--exposure (stack of many frames). 

\section{Detector Performance}
\label{detec}

Although the spectrometer was initially designed to use a subsection 
40x100 of a NICMOS3 detector,
we found later that very good performance can be obtained
on an even larger area. Using 256 pixels in the wavelength direction, we have a
spectral coverage of almost 0.3 $\mu m$. This means that with a single
grating setting we can measure a complete J spectrum and have good coverage
in H and K.

The best 40x256 subsection was selected on the basis of good cosmetics
(low percentage of bad pixels) and low dark current and readout noise. 
We measured the percentage of
bad pixels, the dark current, and the readout noise 
via laboratory tests based on sets of images taken 
at a series of exposure times of a spatially uniformly illuminated
scene, and without any illumination 
(by substituting the filter with a cold stop). 

The readout noise is determined as the mean standard deviation of each pixel 
in the stack of short integration times where the dark current is negligible.
The dark current and gain measurement are based on two linear regressions: 
values of dark frames as a function of exposure time in the first case, 
and spatial 
medians of the stack variance relative to the stack median in the second one.
Details of these tests are presented in
Vanzi et al. (1995). In Table 1, we present the results of further tests 
carried out in April 1995.
\begin{table}
\label{riv}
\begin{center}
\caption{Measured parameters of the detector}
\begin{tabular}{lc}
\hline
Bad pixels   &  2.9\%       \\   
Dark current &  0.9 $e^-$/sec \\
Read out noise &  45 $e^-$    \\
\hline
\end{tabular}
\end{center}
\end{table}

\section{Observations and Data Reduction}
\label{obs}

The procedures for LonGSp observations are those commonly used in
NIR spectroscopy, optimized for the characteristics of the instrument.
For compact sources, observations consist of several
groups of frames with the object placed at different positions
along the slit.
In the case of extended sources, observations consist of several
pairs of object and sky frames.
On-chip integration time is 60 sec
or less for a background level of roughly 6000 counts/pixel because
of ensuing problems with sky line subtraction (see below).
At a given position along
the slit, several frames can be coadded.

The main steps in the reduction of NIR spectroscopic data
are flat--field correction, subtraction of sky emission,
wavelength calibration, correction for telluric absorption, and
correction for optical system $+$ detector efficency.
Data reduction can be performed using the IRSPEC context in MIDAS,
the ESO data analysis package, properly modified to take
into account the LonGSp instrumental setup.
We have found it useful to acquire
dark and flat frames at the beginning and the end of the night;
we obtain flat-field frames by illuminating the dome with a halogen lamp.

Observations of a reference star are taken for a fixed grating position.
An early type, featureless star (preferably an O star) is needed
to correct for telluric absorption and differential
efficency of the system, and a photometric
standard star is needed if one wants to flux calibrate the final spectrum
(only one grating position in each band is required).
An alternative technique, proposed by Maiolino et al (1996),
consists of using a G star corrected through data of solar spectrum.
Both methods have been succesfully tested.

Flat field frames are first corrected for bad pixels, 
then dark-current subtracted,  and
normalized. Dark current is subtracted from all raw frames, and then
divided by the normalized flat field.

For compact sources (the frames taken at the different positions 
along the slit are denoted by A, B and C), 
the sky is subtracted by 
considering $A-B$, $B-(A+C)/2$, and $C-B$, and taking a median of the
three differences.
In case of extended objects, if A and B denote object and sky frames,
the sky is subtracted by considering $(A1+A2)/2-(B1+B2)/2$
(the order of observations is $A1\,B1\,B2\,A2$).
However, a simple sky subtraction is almost never sufficient to
properly eliminate the bright OH lines whose intensity varies
on time scales comparable with object and sky observations.
Moreover, mechanical instabilities can produce movements of spectra
(usually a few hundreds of a pixel) which are nevertheless enough to
produce residuals which exceed the detector noise.
To correct for these two effects, the sky frames are multiplied by a
correcting factor and shifted along the dispersion direction by
a given amount. These factors and shifts are chosen automatically
by minimizing the standard deviation in selected detector areas
where only sky emission is present.
Because this effect increases with the integration time, it is advisable
not to exceed 60 seconds for each single integration.

Slit images at various wavelengths are tilted as a consequence of
the off-axis mount of the grating. Sky subtracted frames
are corrected by computing analytically the tilt angle
from the instrumental calibration parameters, or by directly measuring it
from the data.

Wavelength calibration in LonGSp data is performed using OH sky emission
lines.
The wavelength dispersion on the array is linear to within a small
fraction of the pixel size and is computed analytically
once the central wavelength of the frame is known.
At the beginning of the data reduction,
the nominal central wavelength used in the observations
is assigned to a properly chosen sky frame.
Then the calibration is refined using the bright OH sky lines
(precise wavelengths of OH lines
as well as a discussion of their use as calibrators are given
in Oliva \& Origlia 1992).

The same procedures are applied to the reference stars frames to obtain
the calibration spectra, and the spectrum of a
photometric standard star can be used to flux calibrate the final frames.

\section{Astronomical Results}
\label{astro}

The first tests at the telescope took place successfully in early 1994.
From these observations we measured
the efficiency of the instrument (through the observation of 
photometric standard stars) and its sensitivity (1$\sigma$ in 60 sec of
integration time); these are reported in table 2.
\begin{table}
\label{efi}
\begin{center}
\caption{Astronomical performance}
\begin{tabular}{cccc}
\hline
Band(order) & Efficency & Line$^{1}$ & Continuum$^{2}$ \\
\hline
J (I)    & 0.045  & 4$\times 10^{-14}$  & 2$\times 10^{-15}$  \\
H (I)    & 0.10   & 2$\times 10^{-14}$  & 8$\times 10^{-16}$  \\
K (I)    & 0.08   & 2$\times 10^{-14}$  & 8$\times 10^{-16}$  \\
\hline
\multicolumn{4}{l}{$^1~erg\, cm^{-2}\, s^{-1}$ \hskip 0.4cm 
$^2~erg\, cm^{-2}\, s^{-1}\, {\rm\AA}^{-1}$} \\
\end{tabular}
\end{center}
\end{table}

\begin{figure}
\centerline{\psfig{figure=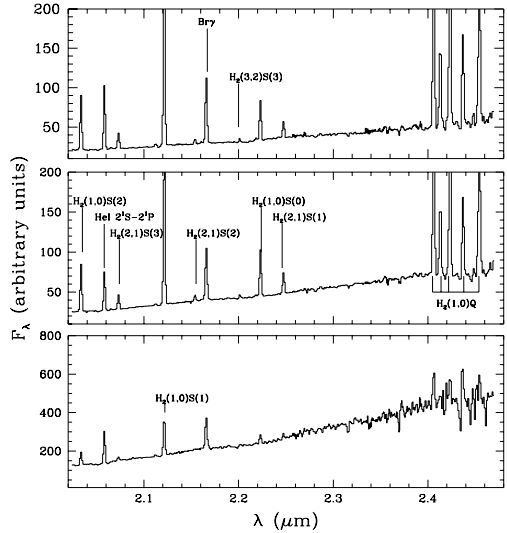,width=7.3cm}}
\caption{ 
Pk2 in Orion, three positions along the slit, with a total integration time
of 60 sec.
}
\end{figure}

\begin{figure}
\centerline{\psfig{figure=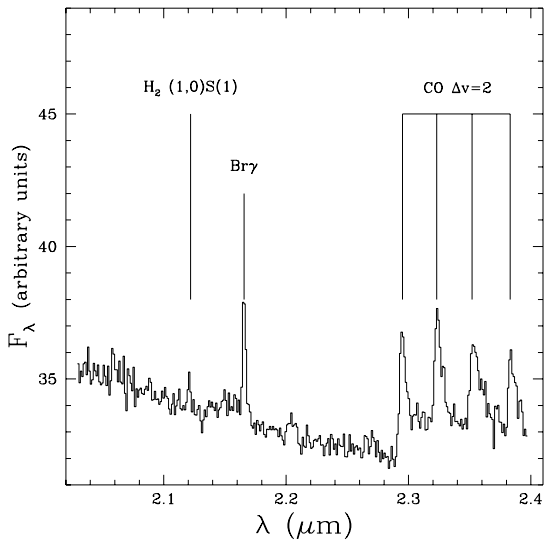,width=7.3cm}}
\caption{ 
The spectrum of LkHa215 (Ae/Be star) in K band, with a total integration time
of 60 sec.
}
\end{figure}

Since January 1996, LonGSp is offered to TIRGO users and employed 
in several galactic and extragalactic programs.
To give an impression of the capabilities of the instrument,
we show (in Figs. 3,4 and 5) some acquired spectra of various type of sources:
extended, compact and extragalactic, without comment as to their astrophysical 
significance.

\begin{figure}
\centerline{\psfig{figure=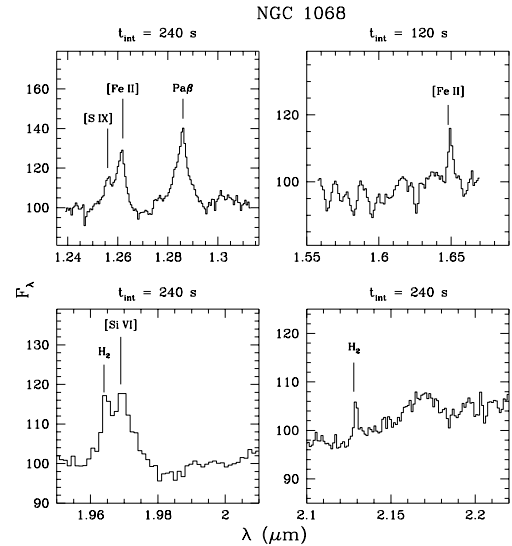,width=7.3cm}}
\caption{ 
Spectrum of the Seyfert 2 NGC 1068. 
}
\end{figure}

\end{document}